# COMPUTATIONAL INTELLIGENCE FOR CONDITION MONITORING

Tshilidzi Marwala
School of Electrical and Information Engineering
University of the Witwatersrand
Private Bag 3, Wits
Johannesburg, 2050, South Africa
e-mail: t.marwala@ee.wits.ac.za
AND
Christina Busisiwe Vilakazi
School of Electrical and Information Engineering
University of the Witwatersrand
Private Bag 3, Wits
Johannesburg, South Africa, 2050
e-mail: busisiwev@yahoo.com

ABSTRACT
Condition monitoring techniques are described in this chapter. Two aspects of condition monitoring process are considered: (1) feature extraction; and (2) condition classification. Feature extraction methods described and implemented are fractals, Kurtosis and Mel-frequency Cepstral Coefficients. Classification methods described and implemented are support vector machines (SVM), hidden Markov models (HMM), Gaussian mixture models (GMM) and extension neural networks (ENN). The effectiveness of these features were tested using SVM, HMM, GMM and ENN on condition monitoring of bearings and are found to give good results.

INTRODUCTION

Condition monitoring of machines is gaining importance in industry due to the need to increase machine reliability and decrease the possible loss of production due to machine breakdown. By definition, condition monitoring is performed when it is necessary to access the state of a machine and to determine whether it is malfunctioning through reason and observation (William et al., 1992). Condition monitoring can also be defined as a technique or process of monitoring the operating characteristics of a machine so that changes and trends of the monitored signal can be used to predict the need for maintenance before a breakdown or serious deterioration occurs, or to estimate the current condition of a machine. Condition monitoring has become increasingly important, for example, in manufacturing companies due to an increase in the need for normal undisturbed operation of equipment in manufacturing. An unexpected fault or shutdown can result in a serious accident and financial loss for a company. Manufacturing companies must find ways to avoid failures, minimize downtime, reduce maintenance costs, and lengthen the lifetime of their equipments. With reliable condition monitoring process, machines can be utilized in a more optimal fashion. Time-based maintenance follows a schedule to decide when maintenance is to be conducted. This leads to inefficiencies because either the maintenance may be conducted needlessly early or a failure may happen before scheduled maintenance takes place. Condition monitoring can therefore be used for condition based maintenance, or predictive maintenance.

Rotating machines are used in various industrial applications. One of the most common components in modern rotating machinery is the rolling element bearing. Most machine failures are linked to bearing failure (Lou and Loparo, 2004), which often result in lengthy downtime that have economic consequences. As a result, an increasing volume of condition monitoring data is captured and presented to engineers. This leads to these key problems: the data volume is too large for engineers to deal with; and the relationship between the plant item, its health and the data generated is not always well understood. Therefore, the extraction of meaningful information from data is difficult. Hence, a reliable, fast and automated diagnostic technique allowing relatively unskilled operators to make important decisions without the need for a condition monitoring specialist to examine the data and diagnose problems is required. The most commonly used condition monitoring system is vibration-based condition monitoring. Vibration monitoring is based on the principle that all systems produce vibration. When a machine is operating properly, vibration is small and constant; however, when faults develop and some of the dynamic processes in the machine change, the vibration spectrum also changes (Marwala, 2001).

The success of a classification systems depend very much on the effectiveness of the extracted features. Another crucial step is to establish a reliable and effective condition monitoring classification system. The objective of this chapter is to give a review of three feature selection techniques that have been used recently for bearing fault diagnosis (Nelwamondo et al., 2006)  These techniques are: 1) Mel-frequency Cepstral Coefficients (MFCC) which is a time-frequency domain analysis technique that has been used extensively in speech recognition; 2) Kurtosis which is the time domain analysis method; 3) Fractal dimension analysis method which is a time domain analysis method that has been applied to problems in image processing.  This chapter also evaluates the effectiveness of the extracted feature for bearing fault diagnosis with the Support Vector Machine (SVM), Hidden Markov Model (HMM), Gaussian Mixture Model (GMM) and Extension Neural Network (ENN) classifiers for bearing fault diagnosis. SVM was chosen as it has been applied successfully in many fault diagnosis applications. The inspiration for the use of GMM and HMM is their success in speech recognition and ENN was chosen because of its success for pattern recognition of partial discharges.

BACKGROUND
As mentioned earlier the success of a classification system depends on the effectiveness of the extracted observation sequence to represent a particular machine state or condition. During the past decades, considerable research effort has been put into the development of various feature extraction techniques and condition monitoring systems. Feature extraction techniques can be classified into three domains, namely; frequency domain analysis, time-frequency domain analysis and time domain analysis (Ericsson et al., 2004). The frequency domain methods often involve frequency analysis of the vibration signals and look at the periodicity of high frequency transients. In the processes, the frequency domain methods search for a train of repetitions occurring at any of the characteristic defect frequencies (Ocak and Loparo, 2004). This procedure gets complicated considering the fact that the periodicity of the signal may be suppressed. These frequency domain techniques include the frequency averaging technique, adaptive noise cancellation and the high frequency resonance technique (HFRT) amongst others. The HFTR is the most popular for bearing fault detection and diagnosis (Ocak and Loparo, 2004). The disadvantage of the HFTR technique is that it requires several impact tests to determine the bearing resonance frequency; hence, it becomes computationally expensive (Ocak and Loparo, 2004). McFadden and Smith (1984) presented the envelope analysis which is another commonly used frequency domain technique for detection and diagnosis of bearing faults. The main disadvantage of the frequency domain analysis is that it tends to average out transient vibrations and therefore becomes more sensitive to background noise. To overcome this problem, the time-frequency domain analysis is used which shows how the frequency contents of the signal changes with time. The examples of such analyses are: Short Time Fourier Transform (STFT), the Wigner-Ville Distribution (WVD) and most commonly the Wavelet Transform (WT). These techniques are studied in detail in (Li et al., 2000). The last category of the feature extraction is the time domain analysis. Time domain methods usually involve indices that are sensitive to impulsive oscillations, such as peak level, root mean square (rms) value, crest factor analysis, kurtosis analysis, shock pulse counting, time series averaging method, signal enveloping method and many more (Ocak and Loparo, 2004; Li et al., 2000). Ericsson *et al.* (2004) and Li *et al.* (2000) showed that unlike the frequency domain analysis, the time-domain analysis is less sensitive to suppressions of the periodicity.

Various feature extraction techniques have been implemented successfully for vibration-based condition monitoring. Loparo (2004; 2005) and Nikolaou (2002) have used wavelet transforms to detect and classify different faults in bearings while Rojas and Nandi (2006) used spectral and statistical features for the classification of bearing faults. Peng (2005) compared the Hilbert-Huang transform with wavelet transform in bearing fault diagnosis. Junsheng (2006) proposed a feature extraction method based on empirical mode decomposition method and autoregressive model for the roller bearings. Baillie and Mathew (1996) implemented an autoregressive modeling that does not only classify, but also provides a one-step-ahead prediction of the vibration signal using the previous outputs. Yang *et al.* (2006) applied the basis pursuit and obtained better results than wavelet transforms. Altman and Mathew (2001) also used the discrete wavelet packet analysis to enhance the detection and diagnostic of rolling element bearing faults. Prabhakar et al. (2002) have also showed that DWT can be used to improve the detection of bearing faults.

The second crucial decision in machine condition monitoring is to choose an effective classification system. Condition classification includes the identification of the operating status of the machine and type of failure by interpreting the representative system condition. The classification system can be classified into two main groups, knowledge-based and data-based models. Knowledge-based models rely on human-like knowledge of the process and its faults. Knowledge-based models like expert systems or decision trees apply human-like knowledge of the process for fault diagnosis. In fault diagnostics, the human expert could be a person who operates the diagnosed machine or process and who is very well aware of different kinds of faults occurring in it. Building the knowledge base can be done by interviewing the human operator on fault occurrences in the diagnosed machine and on their symptoms. Expert systems are usually suitable for problems, where a human expert can linguistically describe the solution. Typical human knowledge is vague and inexact, and handling this kind of information has often been a problem with traditional expert systems. For example, the limit when the temperature in a sauna is too high is vague in human mind. In practice, it is very difficult to obtain adequate representations of the complex and highly non-linear behavior of faulty plants using quantitative models. Knowledge-based models may be utilized together with a simple signal-based diagnostics, if the expert knowledge of the process is available. However, it is often impossible even for a human expert to distinguish faults from the healthy operation, and also multiple information sources may need to be used for trustworthy decision-making. Thus, the data-based models are the most flexible approach to automated condition monitoring.

Data-based models are applied when the process model is not known in the analytical form and expert knowledge of the process performance under faults is not available. The data-based models can be created in numerous ways. During the last years artificial neural network based models like Multilayer Perceptron (MLP) and Radial Basis Function (RBF) have been used extensively for bearing condition monitoring. Samanta *et al.* (2003) used artificial neural network with time-domain features for rolling element bearing detection. Yang *et al*. (2004) applied the ART-KOHONEN to the problem of fault diagnosis of rotating machinery. Lately, kernel-based classifiers such as Support Vector Machine have been used for bearing fault diagnosis. Rojas and Nandi (2006) used SVM for the detection and classification of rolling element bearing faults. Samanta (2004) used both artificial neural networks (ANN) and SVM with genetic algorithm for bearing fault detection. Yang *et al.* (2005) used multi-class SVM for fault diagnosis of rotating machinery. However, data-based statistical approaches have achieved considerable success in speech recognition and have been recently used for condition monitoring. Ertunc *et al*. (2004) used HMM to determine wear status of the drill bits in a drilling process. Ocak and Loparo (2004), Purushotham *et al*. (2005), Miao *et al.* (2006) and Marwala *et al*. (2006) used HMM for bearing fault detection and diagnosis.

FEATURE EXTRACTION
Various features that can be extracted from vibration signals of bearing elements have been investigated. This section presents a brief discussion of fractal analysis, Mel-frequency cepstral coefficients and Kurtosis.

Fractal Dimension
Most of the vibrations are periodic movements with some degree of turbulence. To detect different bearing faults, these non-linear turbulence features must be extracted. The non-linear turbulence features of the vibration signal are quantified using the fractal model (Maragos and Potamianos, 1999). To define the fractal dimension, let the continuous real-valued function, $s(t), 0 \leq t \leq T$ represents a short-time vibration signal. Furthermore, let the compact planar set (Maragos and Potamianos, 1999),

$$F = \{(t, s(t)) \in R^2 : 0 \leq t \leq T\} \tag{1}$$

The fractal dimension of a compact planar set *F* is called the Hausdorff dimension and its value generally lies between 1 and 2 (Maragos and Potamianos, 1999). The problem with this dimension is that it is only a mathematical concept and is therefore extremely hard to compute. Due to this, other methods are used to approximate this dimension such as the Minkowski-Bouligand dimension and Box-Counting dimension (Maragos and Potamianos, 1999). In this study, fractal dimension is approximated using the Box-Counting dimension, which is discussed further in the next section.

*Box-Counting Dimension*

The Box-Counting dimension ($D_B$) of, *F*, is obtained by partitioning the plane with a squares grids of side *ε*, and *N(ε)* number of squares that intersect the plane and is defined as (Falconer, 1952)

$$D_B(F) = \lim_{\varepsilon \to 0} \frac{\ln N(\varepsilon)}{\ln(1/\varepsilon)} \tag{2}$$

Assuming a discrete bearing vibration signal, $s_1, s_2, ..., s_T$ then $D_B$ is given by (Falconer, 1952)

$$D_B(F) = \left\{ J \left( \sum_{j=1}^{J} \ln(1/\varepsilon_j) \cdot \ln(N(\varepsilon)) \right) - \left( \sum_{j=1}^{J} \ln(1/\varepsilon_j) \right) \left( \sum_{j=1}^{J} \ln N(\varepsilon) \right) \right\} / \left\{ J \cdot \sum_{j=1}^{J} (\ln(1/\varepsilon_j))^2 - \left( \sum_{j=1}^{J} \ln(1/\varepsilon) \right)^2 \right\} \tag{3}$$

Where *J* is the computation resolutions and $\varepsilon_{min} \leq \varepsilon_j \leq \varepsilon_{max}$ with $\varepsilon_{max}$ and $\varepsilon_{min}$ represent the maximum and minimum resolutions of computation. In equation 3, $D_B$ is equal to the slope obtained by fitting a line using least squares method (Maragos and Potamianos, 1999).

*Multi-Scale Fractal Dimension (MFD)*

It should be noted that the fractal dimension discussed in the last section is a global measure and therefore does not represent all the fractal characteristics of the vibration signal (Falconer, 1952). To overcome this problem, the Multi-scale fractal dimension set is created. The MFD ($D(s,t)$) of the vibration signal, *s*, is obtained by computing the dimensions over a small time window. This MFD set is obtained by dividing the bearing vibration signal into *K* frames, then *K* maximum computation resolutions are set as (Falconer, 1952):

$$\varepsilon_k^{max} = k.\varepsilon_{min} (1 \leq k \leq K) \tag{4}$$

where $\varepsilon_{min}$ is the same as before, which is the minimum valid resolution of the computation. The Box-Counting dimension expression can then be written as

$$D^k(F) = \left\{ k \left( \sum_{j=1}^{k} \ln(1/j\varepsilon_{min}).\ln(N(j\varepsilon_{min})) \right) - \left( \sum_{j=1}^{k} \ln(1/j\varepsilon_{min}) \right) \left( \sum_{j=1}^{k} \ln N(j\varepsilon_{min}) \right) \right\} / \left\{ k.\sum_{j=1}^{k} (\ln(1/j\varepsilon_{min}))^2 - \left( \sum_{j=1}^{k} \ln(1/j\varepsilon_{min}) \right)^2 \right\} \tag{5}$$

Finally, the corresponding MFD of the vibration signal is given by

$$MFD(s) = \{D^1(s), D^2(s), ...., D^K(s)\} \tag{6}$$

where, $D^k(s)$ is the fractal dimension of the $k^{th}$ frame and this is called the *fractogram* (Maragos and Potamianos, 1999).

Mel-frequency Cepstral Coefficients (MFCCs)
MFCCs have been widely used in the field of speech recognition and are able to represent the dynamic features of a signal as they extract both linear and non-linear properties. MFCC can be a useful tool of feature extraction in vibration signals as vibrations contain both linear and non-linear features. The Mel-frequency Cepstral Coefficients (MFCC) is a type of wavelet in which frequency scales are placed on a linear scale for frequencies less than 1 kHz and on a log scale for frequencies above 1 kHz (Wang et al., 2002). The complex cepstral coefficients obtained from this scale are called the MFCC (Wang et al., 2002). The MFCC contain both time and frequency information of the signal and this makes them useful for feature extraction. The following steps are involved in MFCC computations.

a) Transform input signal, $x(n)$ from time domain to frequency domain by applying Fast Fourier Transform (FFT), using (Wang et al., 2002):

$$Y(m) = \frac{1}{F} \sum_{n=0}^{F-1} x(n)w(n)e^{-j\frac{2\pi}{F}nm} \tag{7}$$

where F is the number of frames, $0 \leq m \leq F-1$ and $w(n)$ is the Hamming window function given by:

$$w(n) = \beta \left( 0.5 - 0.5 \cos \frac{2\pi n}{F-1} \right) \tag{8}$$

where $0 \leq n \leq F-1$ and $\beta$ is the normalization factor defined such that the root mean square of the window is unity (Wang et al., 2002).

b) Mel-frequency wrapping is performed by changing the frequency to the Mel using the following equation (Wang et al., 2002):.

$$mel = 2595 \times \log_{10}(1 + \frac{f_{Hz}}{700}) \qquad (9)$$

Mel-frequency warping uses a filter bank, spaced uniformly on the Mel scale. The filter bank has a triangular band pass frequency response, whose spacing and magnitude are determined by a constant Mel-frequency interval.

c) The final step converts the logarithmic Mel spectrum back to the time domain. The result of this step is what is called the Mel-frequency Cepstral Coefficients. This conversion is achieved by taking the Discrete Cosine Transform of the spectrum as:

$$C_m^i = \sum_{n=0}^{F-1} \cos\left(m\frac{\pi}{F}(n+0.5)\right) \log_{10}(H_n) \qquad (10)$$

Where $0 \leq m \leq L-1$ and $L$ is the number of MFCC extracted form the $i^{th}$ frame of the signal. $H_n$ is the transfer function of the $n^{th}$ filter on the filter bank. These MFCC are then used as a representation of the signal.

Kurtosis
There is a need to deal with the occasional spiking of vibration data, which is caused by some types of faults and to achieve this task Kurtosis is used. Kurtosis features of vibration data have also been successfully used in tool condition monitoring by El-Wardany (1996). The success of Kurtosis in vibration signals is based on the fact that vibration signals of a system under stress or having defects differ from those of a normal system. The sharpness or spiking of the vibration signal changes when there are defects in the system. Kurtosis is a measure of the sharpness of the peak and is defined as the normalized fourth-order central moment of the signal (Wang, 2001). The Kurtosis value is useful in identifying transients and spontaneous events within vibration signals (Wang, 2001) and is one of the accepted criteria in fault detection. The calculated Kurtosis value is typically normalized by the square of the second moment, as shown in equation 11. A high value of Kurtosis implies a sharp distribution peak and indicates that the signal is impulsive in nature (Altman and Matthew, 2002).

$$K = \frac{1}{N} \sum_{i=1}^{N} \frac{(x_i - \bar{x})^4}{\sigma^4} \qquad (11)$$

where $\bar{x}$ the mean and $\sigma$ is the variance.

CLASSIFICATION SYSTEM
Once the relevant features are extracted from vibration signals, these features are used for automatic bearing fault detection and diagnosis by applying them to a non-linear classifier. SVM has outperformed ANN for a number of classification problems especially in condition monitoring. SVM has been successfully used for bearing fault detection and diagnosis (Samanta, 2004). Non-linear classifiers like Gaussian Mixture Model (GMM) and Hidden Markov Model (HMM) have shown to outperform ANN in a number of classification problems but mostly in speech related problems. Only recently, researchers like Purushotham *et al*. (2005) have applied speech pattern classifiers such as HMM to fault detection of mechanical systems due to their success in speech recognition. ENN is a fairly new non-linear classifier and it has shown to outperform ANN for partial discharge pattern recognition (Wang, 2005). A brief background of SVM, GMM, HMM and ENN is presented below.

Support Vector Machine (SVM)
SVM is a powerful widely used technique for solving supervised classification problems due to its generalization ability. In essence, SVM classifiers maximize the margin between the training data and the decision boundary, which can be formulated as a quadratic optimization problem in the feature space. The subsets of patterns that are closest to the decision boundary are called support vectors. For a linearly separable binary classification problem, the construction of a hyperplane, $w^t x + b = 0$, so that the margin between the hyperplane and the nearest point is maximized can be posed as the following quadratic optimization problem

$$\min_{w} \frac{1}{2}(w^T w) \quad (12)$$

subject to,

$$d^i((w^T x^j) + b) \geq 1 \quad (13)$$

where $d^i \in \{-1, 1\}$ stands for $i^{th}$ desired output, $x^i \in R^P$ stands for the $i^{th}$ input sample of the training data set $\{x^i, d^i\}_{i-1}^{N}$. Equation 13 forces a rescaling on *(w,b)* so that the point closest to the hyperplane has a distance of $1/\|w\|$. Maximizing the margin corresponds to minimizing the Euclidean norm of the weight vector. Often in practice, a separating hyperplane does not exist. Hence the constraint is relaxed by introducing slack variable $\xi_{i \geq o,} i = 1, \ldots, N$. The optimization problem for penalty constant *C* becomes

$$\min_{w, \xi} \frac{1}{2}(w^T w) + C \sum_{i=1}^{N} \xi_i \quad (14)$$

subject to

$$d^i((w^T x^j) + b) \geq 1 - \xi_i \quad (15)$$

where,

$$\xi_i \geq 0 \quad with \quad i = 1, \ldots, N$$

The C controls the tradeoff between the robustness of the machine and the number non-separable points. By introducing the Langrange multiplier $\alpha_i$ and using the Karush-Kuhn-Tucker theorem of optimization theory (Bertsekas, 1995), the decision function for the vector $x$, then becomes (Haykins, 1999):

$$f(x) = \text{sgn}\left(\sum_{i=1}^{N} d^i \alpha_i \langle x, x^i \rangle + b\right) \tag{16}$$

By replacing the inner product $\langle x, x^i \rangle = (x^T)(x^i)$ with a kernel function $k(x, x^i)$, the input data are mapped to a higher dimensional space. It is then in this higher dimensional space that a separating hyperplane is constructed to maximize the margin. In the lower dimensional data space, this hyperplane becomes a non-linear separating function. The typical kernel functions are the polynomial kernel $k(x, x^i) = (x \times x^i + 1)^d$ and the Gaussian kernel $k(x, x^i) = \exp(-(x - x^i)^2 / \delta^2)$, where $d$ is the degree of the polynomial and $\delta^2$ is the bandwidth of the Gaussian kernel.

Hidden Markov Model (HMM)
HMM has recently been applied to various conditions monitoring application machine tool monitoring (Owsley et al., 1997), and bearing fault detection (Purushotham et al., 2005; Lou and Loparo, 2004). HMM is a stochastic signal model and is referred to as Markov sources or probabilistic functions of Markov chains (Rabiner, 1989). A Markov chain is a random process of discrete-valued variables that involves a number of states. These states are linked by possible transitions, each with an associated probability and each state has an associated observation. The state transition is only dependent on the current state and not on past states. The actual sequence of states is not observable and hence the name *hidden Markov* (Ertunc et al., 2001). The compact representation for an HMM with a discrete output probability distribution is given by

$$\lambda = \{A, B, \pi\} \tag{17}$$

Where $\lambda$ is the model, $A = \{a_{ij}\}$, $B = \{b_{ij}(k)\}$ and $\pi = \{\pi_i\}$ is a transition probability distribution, the observation probability distribution and initial state distribution, respectively. These parameters of a given state, $S_i$, are defined as (Rabiner, 1989)

$$a_{ij} = P(q_{t+1} = S_j \mid q_t = S_i), \quad 1 \leq i, j \leq N \tag{18}$$

$$b_{ij}(k) = P(o_k \mid q_t = S_i), \quad 1 \leq j \leq N, 1 \leq k \leq M \tag{19}$$

and

$$\pi_i = P(q_1 = S_i)), \quad 1 \leq i \leq N \tag{20}$$

where $q_t$ the state at time $t$ and $N$ is denotes the number of states. Furthermore, $o_k$ is the $k^{th}$ observation and $M$ is the number of distinct observation. HMM is thus a finite-state machine which changes state every time unit. There are three basic HMM problems to be solved. Firstly, evaluation, which finds the probability of the observation sequence, $O = o_1, o_2, ..., o_T$, of visible states generated by the model $\lambda$. Using the model in Equation 21, the probability is computed as (Rabiner, 1989):

$$P(O, \lambda) = \sum_{allS} \pi_{S_0} \prod_{t=0}^{T=1} a_{S_t S_{t+1}} b_{S_{t+1}}(o_{S_{t+1}}) \tag{21}$$

Secondly, decoding, which finds a state sequence that maximizes the probability of observation sequence is realized by the so-called Viterbi algorithm (Viterbi, 1967). Lastly, training which adjusts model parameters to maximize probability of observed sequence.

Gaussian Mixture Models (GMM)
Gaussian mixture models have been a reliable classification tool in many applications of pattern recognition, particularly in speech and face recognition. GMM have proved to perform better than Hidden Markov Models in text independent speaker recognition (Reynolds et al., 2003). The success of GMM in classification of dynamic signals has also been demonstrated by many researchers such as Cardinaux and Marcel (2003) who compared GMM and MLP in face recognition and found that the GMM approach easily outperforms the MLP approach for high resolution faces and is significantly more robust in imperfectly located faces. Other advantages of using GMM are that it is computationally inexpensive and is based on well understood statistical models (Reynolds et al., 2003). GMM works by creating a model of each fault which is written as (Reynolds et al., 2003):

$$\lambda = (w, \mu, \Sigma) \tag{22}$$

where $\lambda$ is the model, $w$ represents the weights assigned to the Gaussian means, $\mu$ is the diagonal variance of the features used to model the fault and $\Sigma$ is the covariance matrix. GMM contains a probability density function of the observation consisting of a sum of normal observations. A weighted sum of Gaussians normally provides an accurate model of the data. Each Gaussian comprises a mean and a covariance, hence, a mixture of components. The Gaussian probability density function is given by (Reynolds et al., 2003):

$$p(x) = \frac{1}{\sigma\sqrt{2x}} e^{\frac{-(x-\mu)^2}{2\sigma^2}} \tag{23}$$

where $\mu$ is the mean and $\sigma$ is the standard deviation of the distribution of a variable $x$. For a case where $x$ is a vector of features, equation (23) becomes (Reynolds et al., 2003)

$$p(x) = \frac{1}{\sqrt{(2\pi)^n |\Sigma|}} e^{-\frac{1}{2}\left[(x-\mu)'\Sigma^{-1}(x-\mu)\right]} \qquad (24)$$

where $n$ is the size of the vector feature, $x$. The log-likelihood is then computed as (Reynolds et al., 2003)

$$\hat{s} = \arg\max_{1 \leq f \leq F} \sum_{k=1}^{K} \log p(x_k | \lambda_f) \qquad (25)$$

where $f$ represents the index of the type of fault, whereas $F$ is the total number of known fault conditions and $x = \{x_1, x_2, ..., x_K\}$ is the unknown fault vibration segment. $P(x_k|\lambda_f)$ is the mixture density function. An arbitrary probability density of a sample vector $x$ can be approximated by a mixture of Gaussian densities (Bishop, 1995) as

$$p(x | \lambda) = \sum_{i=1}^{M} w_i p(x) \qquad (26)$$

where all mixtures weights, $w_i$, are adjusted to satisfy the constrains (Meng et al., 2005) and $0 \leq w_i \leq 1$. Training GMM is a fast and straightforward process which estimates the mean and covariance parameters from the training data (Meng et al., 2005). The training procedure estimates the model parameters from a set of observations using the Expectation Maximization (EM) algorithm. The EM algorithm tries to increase the expected log-likelihood of the complete data $x$ given the partially observed data (Meng et al., 2005) and finds the optimum model parameters by iteratively refining GMM parameters for a given bearing fault feature vector.

Extension Neural Network (ENN)
ENN is a new pattern classification system based on concepts from neural networks and extension theory (Wang and Hung, 2003). The extension theory uses a novel distance measurement for classification processes, and the neural network can embed the salient features of parallel computation and learning capability. The classifier is well suited to classification problems where there exist patterns with a wide range of continuous inputs and a discrete output indicating which class the pattern belongs to. ENN comprises of an input layer and an output layer. The input layer nodes receive an input feature pattern and use a set of weighted parameters to generate an image of the input pattern. There are two connection weights between input nodes and output nodes; one connection represents the lower bound for this classical domain of features and the other represents the upper bound. The entire network is thus represented by a matrix of weights for the upper and lower limits of the features for each class $W_U$ and $W_L$. A third matrix representing the cluster centers is also defined as:

$$z = \frac{W_u + W_l}{2} \tag{27}$$

ENN uses supervised learning, which tunes the weights of the ENN to achieve a good clustering performance or to minimize the clustering error. The network is trained by adjusting the network weights and recalculating the network centers for each training pattern depending on the extension distance (ED) of that pattern to its labeled cluster. Each training pattern adjusts the network weights and the centers by amounts that depend on the learning rate. In general, the weight update for a variable $x_i$ is:

$$w^{new} = w^{old} - \eta(x_i - w^{old}) \tag{28}$$

where $\eta$ is the learning rate and $w$ can either be the upper or the lower weight matrices of the network centers. It can be shown that for $t$ training patterns for a particular class $C$, the weight is given by (Mohamed et al., 2006):

$$w^c(t) = (1-\eta)w^c(0) - \eta \sum (1-\eta)^{t-1} x_i^c \tag{29}$$

This equation demonstrates how each training pattern reinforces the learning in the network by having the most recent signal determines only a fraction of the current value of $z_c^t$. The equation indicates that there is no convergence of the weight values since the learning process is adaptive and reinforcing. Equation 29 also highlights the importance of the learning rate, $\eta$. Small values of $\eta$ require many training epochs, whereas large values may results in oscillatory behavior of the network weights, resulting in poor classification performance.

PROPOSED ARCHITECTURE
The architecture of the proposed framework is shown in Figure 1. As shown in this figure the framework consists of two major stages, namely, data pre-processing with feature extraction and classification stage.

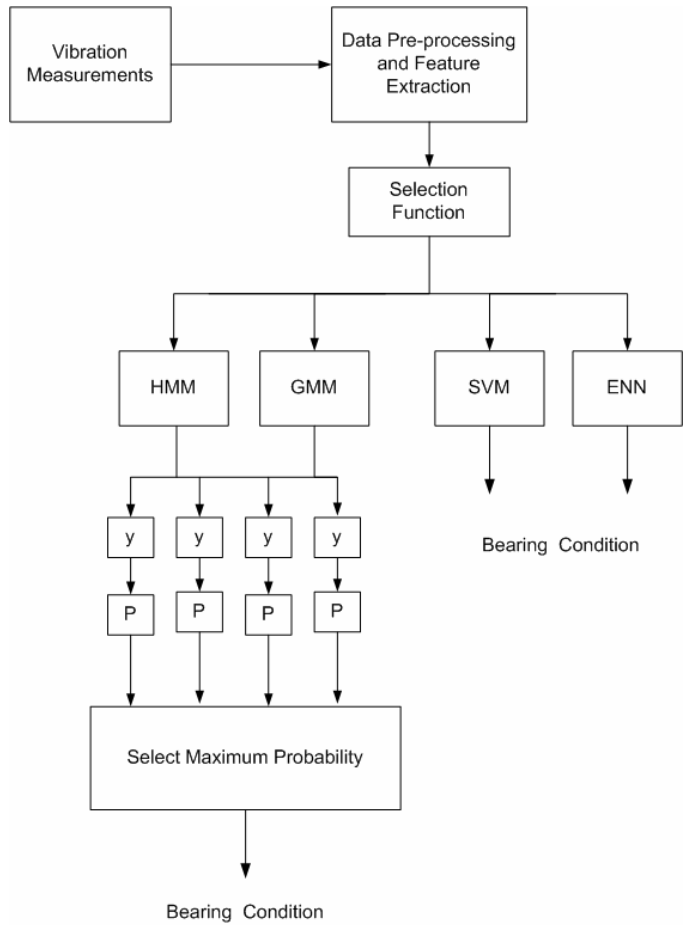

Figure 1: The proposed block diagram for the condition monitoring

The vibration signal is preprocessed by firstly dividing the vibration signals into segments of equal lengths. Preprocessing is followed by extraction of features of each window using feature extraction techniques as explained earlier. The vibration signal was first broken into segments, each being five revolutions long. Firstly, MFCC features were extracted from each frame of the signal. During the extraction of MFCC features, each segment of the signal was further broken into 14 frames of equal duration. The number of MFCC features was varied from 9 to 16 in order to obtain the optimal value. Secondly, fractal features were extracted in the same format with different MFD size. The MFD size was also varied from 2 to 20 in order to obtain the MFD size that gives best results. Lastly, one Kurtosis feature from each segment was extracted. When relevant features are extracted, reference models that will be used to classify faults are built and are used to classify the fault conditions. SVM, GMM, HMM and ENN classifiers were implemented. Figure 1 also shows that both GMM and HMM build models for all possible machine condition and the normal condition. For GMM and HMM, diagnosis of the bearing fault is achieved by calculating the probability of the feature vector, given the entire previously constructed fault model. GMM or HMM with maximum probability then determines the bearing condition.

RESULTS AND DISCUSSION

Vibration Data

The investigation in this chapter is based on the data obtained from Case Western Reserve University website (Loparo, 1998). The experimental setup is comprised of a Reliance Electric 2HP IQPreAlert connected to a dynamometer. Faults of size 0.007, 0.014, 0.021 and 0.028 inches were introduced into the drive-end bearing of a motor using the Electric Discharge Machining (EDM) method. These faults were introduced separately at the inner raceway, rolling element and outer raceway. An impulsive force was applied to the motor shaft and the resulting vibration was measured using two accelerometers, one mounted on the motor housing and the other on the outer race of the drive-end bearing. All signals were recorded at a sampling frequency of 12 kHz.

Results

Figure 2 shows samples bearing vibration signals for the four bearing conditions. In order to diagnose faults, features need to be extracted and be classified.

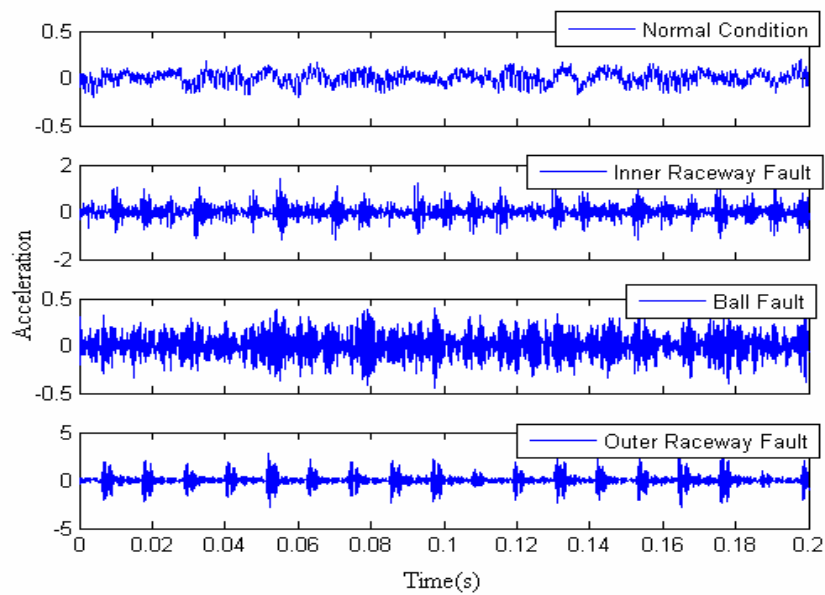

Figure 2: Vibration signals on the bearing under normal condition, inner race fault, outer race fault and ball fault conditions

The optimal classifier parameters were found using exhaustive search. The optimum SVM architecture used polynomial kernel function with degree of 5. The optimum HMM architecture used for experimentation is a 2 state model with diagonal covariance matrix that contains 10 Gaussian mixtures. GMM architecture also used diagonal covariance matrix with 3 centers. The main advantage of using the diagonal covariance matrix for GMM and HMM is that this de-correlates the feature vectors. This is necessary since fractal dimensions are highly correlated values. ENN architecture with an optimal learning rate of 0.219 was used.

The first set of experiments measures the effectiveness of the time domain fractal dimension based feature extraction using vibration signal. Figure 3 shows the MFD feature vector which extracts the bearing fault specific information, plotted for the first second of the vibration signal. Figure 3 shows that the proposed feature extraction technique does extract fault specific features which are used to classify different bearing conditions. However, the optimum size of the MFD must be found.

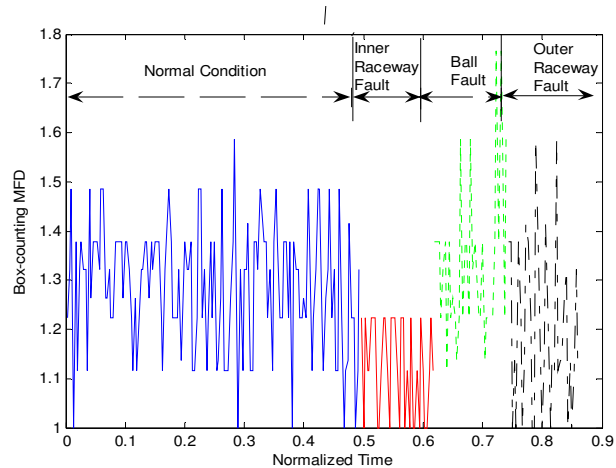

Figure 3: MFD feature extraction comparison for the normal, inner, outer and ball fault for the 1s vibration signal

Figure 4 shows the change of the system accuracy with the change of the MFD size. This figure shows that the size of MFD does not affect the classification accuracy of SVM and ENN. It also shows that the GMM generally has a large optimum MFD size of 13 compared to 5 for HMM.

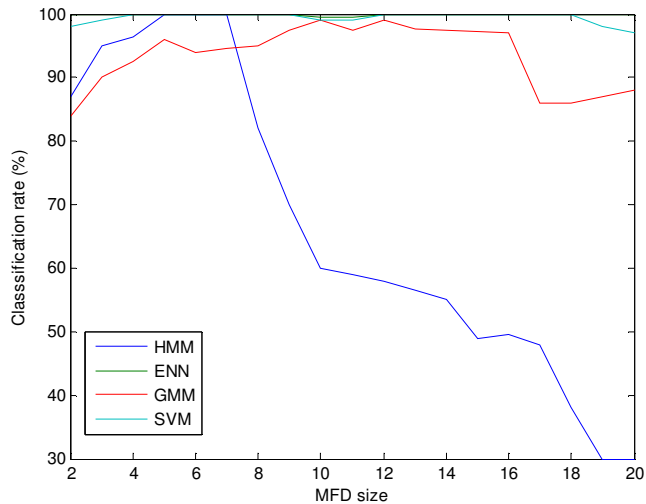

Figure 4: The graph of the change classification rate with change in MFD size

Using optimum SVM, HMM, GMM and ENN architecture together with MFD, the confusion matrix obtained for different bearing faults are presented for HMM, GMM and ENN in Table 1.

Table 1: The confusion matrix for the SVM, HMM, GMM and ENN classifier used with Fractal features

|        | SVM    |       |       |      | HMM    |       |       |      |
|--------|--------|-------|-------|------|--------|-------|-------|------|
|        | Normal | Inner | Outer | Ball | Normal | Inner | Outer | Ball |
| Normal | 100    | 0     | 0     | 0    | 100    | 0     | 0     | 0    |
| Inner  | 0      | 100   | 0     | 0    | 0      | 100   | 0     | 0    |
| Outer  | 0      | 0     | 100   | 0    | 0      | 0     | 100   | 0    |
| Ball   | 0      | 0     | 0     | 100  | 0      | 0     | 0     | 100  |
|        | GMM    |       |       |      | ENN    |       |       |      |
|        | Normal | Inner | Outer | Ball | Normal | Inner | Outer | Ball |
| Normal | 100    | 0     | 0     | 0    | 100    | 0     | 0     | 0    |
| Inner  | 0      | 100   | 0     | 0    | 0      | 100   | 0     | 0    |
| Outer  | 0      | 0     | 100   | 0    | 0      | 0     | 100   | 0    |
| Ball   | 1.8    | 0     | 0     | 98.2 | 0      | 0     | 0     | 100  |

We further investigate the use of MFCC with the GMM, HMM and ENN. We first observe the effect of varying the number of MFCC on the classification rate for the classifiers.

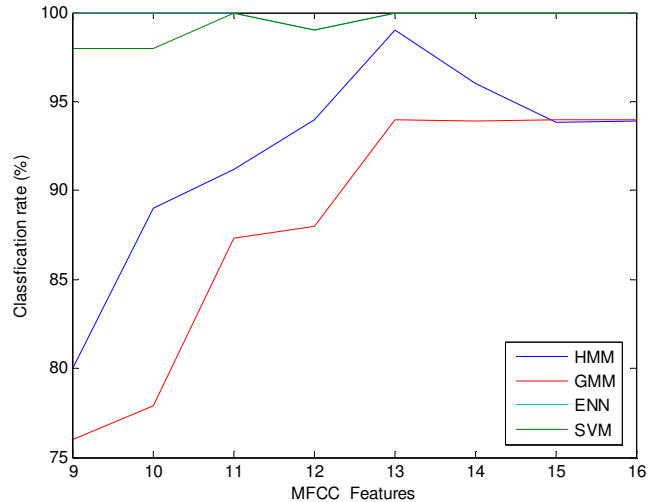

Figure 5: Change in classification rate with change in the number of MFCC

The figure shows that the number of MFCC does not affect the classification accuracy of SVM and ENN. Figure 5 shows that 13 MFCC give optimal results for both HMM and GMM and will therefore be used. It is observed that increasing the number of MFCC above 13 does not improve the classification results.

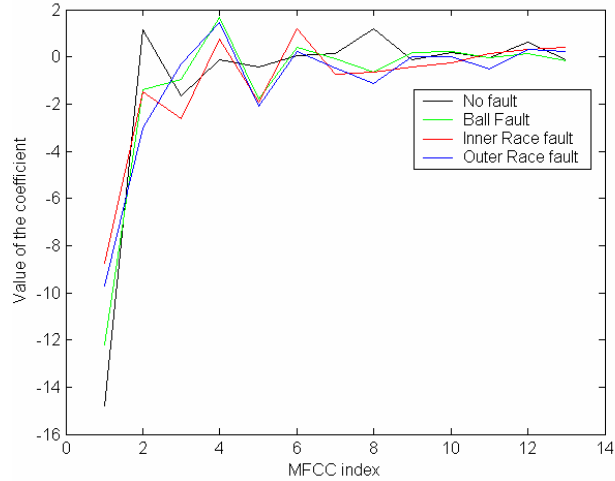

Figure 6: MFCC values corresponding to different fault conditions

The results obtained with MFCC features are shown in the confusion matrix in Table 2. On trying to improve the performance of the MFCC with GMM, we now add the Kurtosis features to the MFCC. The Kurtosis was computed in time for each segment of the signal. Figure 6 shows how Kurtosis of the vibration signals varies with different fault conditions for 60 segments of the vibration data.

Table 2: The confusion matrix for the SVM, GMM, HMM and ENN classifiers used with MFCC features

|  | SVM | | | | HMM | | | |
|---|---|---|---|---|---|---|---|---|
|  | Normal | Inner | Outer | Ball | Normal | Inner | Outer | Ball |
| Normal | 100 | 0 | 0 | 0 | 100 | 0 | 0 | 0 |
| Inner | 0 | 100 | 0 | 0 | 0 | 100 | 0 | 0 |
| Outer | 0 | 0 | 100 | 0 | 0 | 0 | 100 | 0 |
| Ball | 0 | 0 | 0 | 100 | 0 | 0 | 0 | 100 |
|  | GMM | | | | ENN | | | |
|  | Normal | Inner | Outer | Ball | Normal | Inner | Outer | Ball |
| Normal | 86.6 | 0 | 0 | 13 | 100 | 0 | 0 | 0 |
| Inner | 0 | 96.6 | 3.3 | 0 | 0 | 100 | 0 | 0 |
| Outer | 0 | 0 | 100 | 0 | 0 | 0 | 100 | 0 |
| Ball | 3.7 | 1.8 | 0 | 94 | 0 | 0 | 0 | 100 |

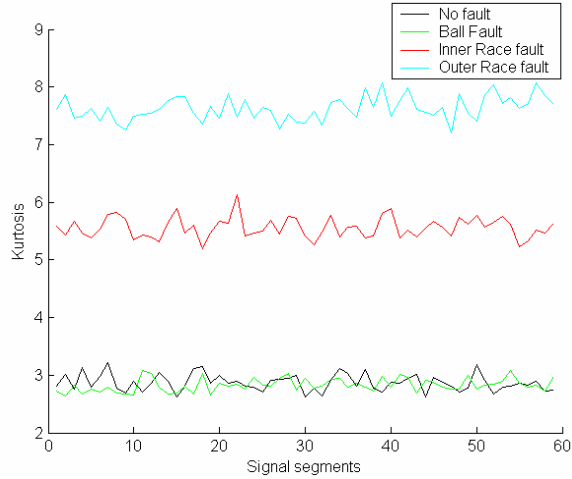

**Figure 7:** Kurtosis values corresponding to different fault conditions

When Kurtosis values are added to the MFCC, an improvement of about 5% is obtained for the GMM. The good classification accuracy of classifier is due to the fact that features seem to be more unique for different fault conditions. Overall results are summarized in Table 3.

Table 3: Summary of classification results

|  | *SVM (%)* | *HMM (%)* | *GMM (%)* | *ENN (%)* |
|---|---|---|---|---|
| Fractal Features | 100 | 100 | 99 | 100 |
| MFCC | 100 | 99 | 94 | 100 |
| MFCC + Kurtosis | 100 | 100 | 99 | 100 |

CONCLUSION

The chapter discussed the two crucial requirements for an automated condition monitoring system. The first requirement is a feature extraction technique that can effectively and accurately extract the condition specific features. The second requirement is a classification system that can effectively classify the machine conditions. This chapter gives a review of three feature extraction techniques that are used for condition monitoring. These techniques are fractal analysis, Mel Frequency Cepstral Coefficients and kurtosis. The effectiveness of the extracted feature was tested using four classifiers, namely, support vector machines, hidden Markov models, Gaussian mixture models and the extension neural network. As vibration-based condition monitoring is the most popular condition monitoring system in machine, the proposed condition monitoring system is tested on bearings. The proposed system gives very good results for bearing fault diagnosis. It should be noted, however, that the proposed condition monitoring system can be applied to various condition monitoring.

*Future Research Direction*

The proposed system gave very good results for bearing condition monitoring, and it should be noted that this system is not limited to bearings only. The general trend is that condition monitoring systems require knowledge and understanding of the links between actual faults and the data generated as a result of these faults. When such understanding does not exist, often detailed data mining and analysis activities are undertaken to derive the required knowledge, data patterns, and relationships. Due to the need for automated condition monitoring system, the latter approach has been applied extensively in various condition monitoring systems.  However, many new condition monitoring systems are implemented when there is no underlying knowledge of the data expected. Neither is there any historical data from which patterns and trends can be derived. These systems use online learning algorithm that enable them to incrementally learn the system as data becomes available.  Furthermore, it is now widely recognized that problems due to the complexity of condition monitoring can be overcome with architectures that contain many dynamically interacting intelligent distributed modules, called intelligent agents.